\documentclass{article}

\usepackage[preprint]{neurips_2026} 

\usepackage[utf8]{inputenc} 
\usepackage[T1]{fontenc}    
\usepackage{hyperref}       
\usepackage{url}            
\usepackage{booktabs}       
\usepackage{amsfonts}       
\usepackage{nicefrac}       
\usepackage{microtype}      
\usepackage{xcolor}         

\usepackage[english]{babel}
\usepackage{graphicx}
\usepackage{algorithm} 
\usepackage{algorithmic}
\usepackage{amsmath}
\usepackage{float}
\usepackage{caption}   
\usepackage{multirow}  
\usepackage{listings}

\lstdefinelanguage{YAML}{
  keywords={true,false,null,y,n},
  keywordstyle=\color{darkgray}\bfseries,
  basicstyle=\ttfamily\footnotesize,
  comment=[l]{\#},
  morestring=[b]',
  morestring=[b]"
}
\title{MPC-Patch-Bench: Security-Aware LLM Code Patch for Multi-Party Computation}

\author{%
Yukuan Zhang\quad Mengxin Zheng \quad Qian Lou \\
University of Central Florida \\
\texttt{\{yukuan.zhang, mengxin.zheng, qian.lou\}@ucf.edu}
}
\begin{document}

\maketitle

\begin{abstract}
Repository-level benchmarks for evaluating Large Language Model (LLM) code repair on Secure Multi-Party Computation (MPC) software do not yet exist, and directly transplanting general-purpose benchmarks such as SWE-bench fails on three structural fronts: (i) MPC repositories are dominated by generic Python infrastructure rather than cryptographic logic; (ii) high-value MPC fixes lack the standardized tests rigid extraction pipelines require; and (iii) standard fail-to-pass evaluation is insufficient for code that must also be cryptographically safe. MPC is increasingly deployed for privacy-preserving machine learning, biomedical collaboration, and secure analytics. Existing MPC-specific code-synthesis efforts cover only operator-level or single-framework tasks; evaluating LLM agents on real repository-level MPC repair instead demands MPC-aware data curation and a verifier matched to the security and numerical-fidelity guarantees MPC programs must obey---neither of which existing benchmarks provide. We introduce \textbf{MPC-Patch-Bench}, a repository-level benchmark organised around two frameworks. \textbf{(1)~The Data Curation Framework} combines a \emph{domain-specific curation agent} that filters raw pull requests through three cryptographic layers with a \emph{human-AI completion engine} that synthesizes missing problem statements and Fail-to-Pass/Pass-to-Pass tests, yielding 205 fully verified instances. \textbf{(2)~The MPC Verifier} provides dedicated security and numerical-fidelity checks via dynamic differential testing against plaintext oracles and MPC-specific static analysis rules that flag unsafe reveals, insecure arithmetic, and illegal public/private casts. The strongest evaluated LLM functionally resolves only 22.9\% of MPC-Patch-Bench tasks; the MPC Verifier further reduces verified resolution to 17.1\%, with up to 40\% of functionally-passing patches rejected for cryptographic or numerical-fidelity violations. These results demonstrate that MPC repair demands capabilities beyond what general-purpose benchmarks capture, and that functional correctness alone substantially overestimates LLM reliability in privacy-preserving software.
\end{abstract}

\section{Introduction}
\label{sec:intro}

Secure Multi-Party Computation (MPC) enables multiple parties to jointly compute a function over private inputs without revealing those inputs to one another. This capability is increasingly important in settings where organizations need to collaborate but cannot directly share raw data, including privacy-preserving machine learning (PPML) \cite{mohassel2017secureml,wagh2019securenn,zheng2019helen}, biomedical collaboration \cite{hie2018realizing}, and secure governmental or institutional analytics \cite{bogdanov2016students}. In contrast to ordinary plaintext computation, MPC executes programs over secret-shared values, interactive protocols, finite rings, and fixed-point encodings. This makes MPC software a critical but unusually demanding target for AI-assisted programming.

Building a repository-level benchmark for LLM-driven MPC code repair is structurally difficult. Naive SWE-bench~\cite{jimenez2024swebench} transplant fails on three fronts: (1) MPC repositories are dominated by generic infrastructure rather than cryptographic logic; (2) MPC fixes frequently ship without standardized tests; and (3) fail-to-pass evaluation cannot detect MPC-specific security or numerical-fidelity violations. The first front is illustrated by the typical content of a merged MPC PR: most updates touch CI configuration or high-level API wrappers rather than protocol logic, so a benchmark drawn naively from these repositories measures ordinary software engineering inside MPC projects. The second is empirically severe---PySyft, our largest source, links only 6.2\% of PRs to issues, and strict SWE-bench extraction retains only 42 instances out of 7{,}305 raw MPC PRs (Appendix~\ref{app:strict_swe_filtering}). The third is rooted in the correctness contract MPC programs must obey: an MPC patch must preserve secrecy across secret shares, respect data-oblivious control flow and public/private type discipline, and avoid finite-ring or fixed-point drift \cite{rastogi2014wysteria,zahur2015oblivc,sok_mpc}---none of which fail-to-pass tests can detect.

\begin{figure}[htbp]
  \centering
  \includegraphics[width=0.85\textwidth]{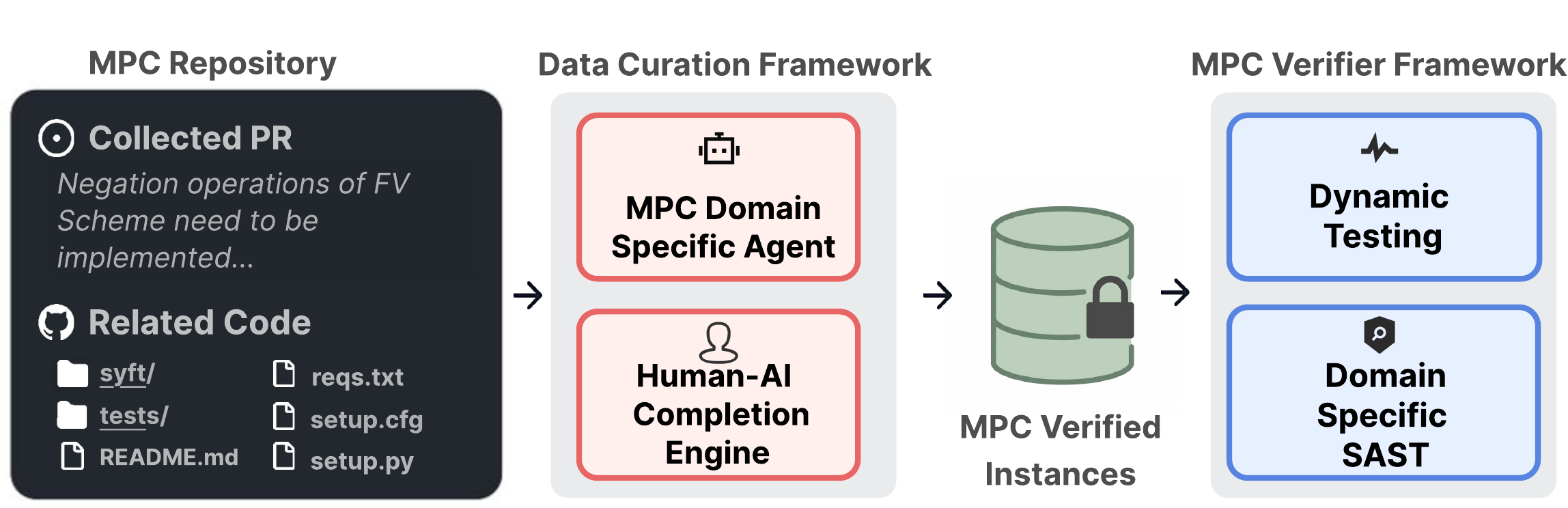}
  \caption{Overview of MPC-Patch-Bench: the Data Curation Framework produces the verified instance pool, which is then evaluated by the MPC Verifier.}
  \label{fig:main_image}
\end{figure}

To address these gaps, we introduce \textbf{MPC-Patch-Bench}\footnote{Anonymous code and dataset for review: \url{https://anonymous.4open.science/r/MPC_bench-D496/}.} (Figure~\ref{fig:main_image}), a repository-level benchmark for evaluating LLMs on real-world MPC code repair.

The \textbf{Data Curation Framework} (Figure~\ref{fig:curation_pipeline}) combines a \emph{domain-specific curation agent} that retains a PR only when the repair touches MPC-specific semantics---cryptographic primitives, multi-party protocol logic, or constraints absent from ordinary Python (data-obliviousness, finite-ring truncation, public/private type discipline)---with a \emph{human-AI completion engine} that uses a human-validated synthesis process to rescue high-value MPC fixes arriving without standardized fail-to-pass or pass-to-pass tests, producing executable benchmark instances rather than discarding them. The \textbf{MPC Verifier} (Figure~\ref{fig:verifier_pipeline}) augments functional tests with a dynamic stream that compares secure outputs against plaintext reference behaviour to catch numerical-fidelity drift \cite{pang2024mpcdiff}, and a static stream that flags MPC anti-patterns such as unsafe reveals, insecure arithmetic, and illegal public/private casts.

\noindent The Data Curation Framework reduces 7{,}305 raw MPC pull requests to 1{,}175 cryptographically relevant candidates and then synthesises missing problem statements and tests, yielding 205 fully verified executable instances. Across eight evaluated LLMs the strongest functionally resolves only 22.9\% of these tasks, dropping to 17.1\% under the MPC Verifier with up to 40\% of functionally-passing patches rejected. Beyond MPC, these results suggest that functional code-repair benchmarks systematically overestimate LLM reliability in security-sensitive software, motivating a class of evaluations matched to the safety guarantees the underlying systems must provide.

\section{Background and Related Work}
\label{sec:related}

\subsection{Secure Multi-Party Computation}
\label{sec:bg_mpc}

Building on the introduction, we briefly fix terminology before discussing related work. Practical MPC implementations split each input into additive secret shares over a finite ring; parties evaluate linear operations locally and realise non-linear operations through interactive sub-protocols built on Beaver triples and oblivious transfer \cite{sok_mpc}. Composed, these primitives support secure pipelines from linear regression to deep neural-network inference, and have driven the emergence of a mature open-source ecosystem.

\noindent
\begin{minipage}[t]{0.62\textwidth}
    \vspace{0pt}
    Our benchmark draws repository-level tasks from five widely-used MPC frameworks. \textbf{CrypTen}~\cite{knott2021crypten} provides a PyTorch-style API for secure tensor computation focused on privacy-preserving machine learning. \textbf{tf-encrypted}~\cite{dahl2018tfencrypted} offers an analogous TensorFlow-native interface for encrypted ML. \textbf{MP-SPDZ}~\cite{keller2020mp} is a protocol-level toolkit supporting dozens of MPC protocols (semi-honest and malicious, two-party and multi-party) compiled from a domain-specific language. \textbf{SecretFlow}~\cite{ma2023secretflow} integrates MPC with federated and hybrid privacy-preserving computation across heterogeneous data silos. \textbf{PySyft}~\cite{ziller2021pysyft} is the OpenMined privacy-preserving deep-learning framework, supporting federated learning, differential privacy, and secure aggregation alongside secret-sharing-based MPC.
\end{minipage}
\hfill
\begin{minipage}[t]{0.33\textwidth}
    \vspace{0pt}
    \centering
    \small
    \makeatletter\def\@captype{table}\makeatother
    \caption{Raw PR statistics.}
    \label{tab:raw-data}
    \begin{tabular}{lr}
        \toprule
        \textbf{Repository} & \textbf{Total PRs} \\
        \midrule
        crypten      & 236   \\
        secretflow   & 658   \\
        MP-SPDZ      & 110   \\
        PySyft       & 5,840 \\
        tf-encrypted & 461   \\
        \midrule
        \textbf{Total} & \textbf{7,305} \\
        \bottomrule
    \end{tabular}
\end{minipage}

\subsection{MPC-Specific Code Synthesis and Verification}

Existing efforts targeting LLM-based MPC code synthesis or MPC-aware verification address slices of the MPC software stack rather than repository-level repair. SPDZCoder \cite{dong2024spdzcoder} combines expert-derived transformation rules with an LLM to translate high-level Python into the MP-SPDZ \cite{keller2020mp} domain-specific language without dedicated training data. While SPDZCoder reduces the manual effort of writing MPC code, it differs from MPC-Patch-Bench in two respects: it operates at the operator/protocol level rather than across full repository contexts, and it targets a single MPC framework (MP-SPDZ), whereas real MPC engineering spans a heterogeneous library ecosystem. CrypTorch \cite{liu2025cryptorch} is an auto-tuning compiler that selects operator approximations for MPC-based ML workloads to balance performance and accuracy; like SPDZCoder, it is a code-generation tool rather than an LLM evaluation, and its objective is performance-vs-accuracy rather than security correctness.

For verification, MPCDiff \cite{pang2024mpcdiff} performs differential testing between secure and plaintext executions to surface numerical-precision deviations in MPC-hardened ML models, HawkEye \cite{ruan2025hawkeye} statically profiles the communication cost of models in multi-party-learning frameworks, and Li et al.\ \cite{li2024metamorphic} use metamorphic testing to uncover logic vulnerabilities in MPC compilers. These tools are complementary to ours rather than competing: MPC-Patch-Bench's verification framework integrates a differential-testing stream similar in spirit to MPCDiff, and a static-analysis stream that targets a deliberately broad set of MPC anti-patterns---unsafe reveals, illegal public/private casts, and insecure randomness---and applies them as a benchmark verifier on LLM-generated patches rather than as standalone tools on hand-written code or compiled protocols.

\subsection{General-Purpose Code Repair Benchmarks}

LLM coding benchmarks have evolved from synthesising isolated functions (HumanEval \cite{chen2021evaluatinglargelanguagemodels}, MBPP \cite{austin2021mbpp}) toward repository-level repair. EvalPlus \cite{liu2023evalplus} and DS-1000 \cite{lai2023ds1000} strengthened test rigour and broadened domain coverage to data-science libraries; CoderEval \cite{yu2024codereval} extended evaluation to function-level generation under realistic repository dependencies. The current standard for repository-level evaluation is SWE-bench \cite{jimenez2024swebench}, which assembles real GitHub issues paired with developer-authored fail-to-pass tests, marking an instance resolved only when the model's patch passes those tests while preserving pass-to-pass behaviour. SWE-smith \cite{yang2025swesmith} extends this paradigm by synthetically generating bug-fix instances at scale via test-breaking mutations on existing codebases.

While SWE-bench captures essential aspects of real-world software repair, it does not fit MPC repositories: its instance selection is domain-agnostic, its strict pre-existing-test requirement reduces 7{,}305 raw MPC PRs to only 42 (Appendix~\ref{app:strict_swe_filtering}), and its purely functional resolution criterion admits patches that leak secrets or branch on private values---exactly the failure modes our MPC Verifier targets.

\section{MPC-Patch-Bench}
\subsection{Overview}
\label{sec:mpc_bench_overview}

MPC-Patch-Bench is built by a multi-stage pipeline that converts raw open-source pull requests into security-aware benchmark instances, embodying two methodological frameworks. The \textbf{Data Curation Framework} (Section~\ref{sec:data_curation}) combines a domain-specific curation agent that filters candidates through three cryptographic layers, retaining only PRs that exercise MPC reasoning, with a human-AI completion engine that rescues high-value but incomplete PRs by co-piloting with a human expert to synthesize problem statements, F2P/P2P tests, and gold patches into executable instances; together these two stages distill 7{,}305 raw MPC pull requests into 205 fully verified instances. The \textbf{MPC Verifier} (Section~\ref{sec:verification_framework}) then augments standard functional tests with a dynamic differential-checking stream and an MPC-specific static-analysis stream, exposing failure modes invisible to ordinary unit tests.

\begin{figure*}[htbp]
    \centering
    \includegraphics[width=\textwidth]{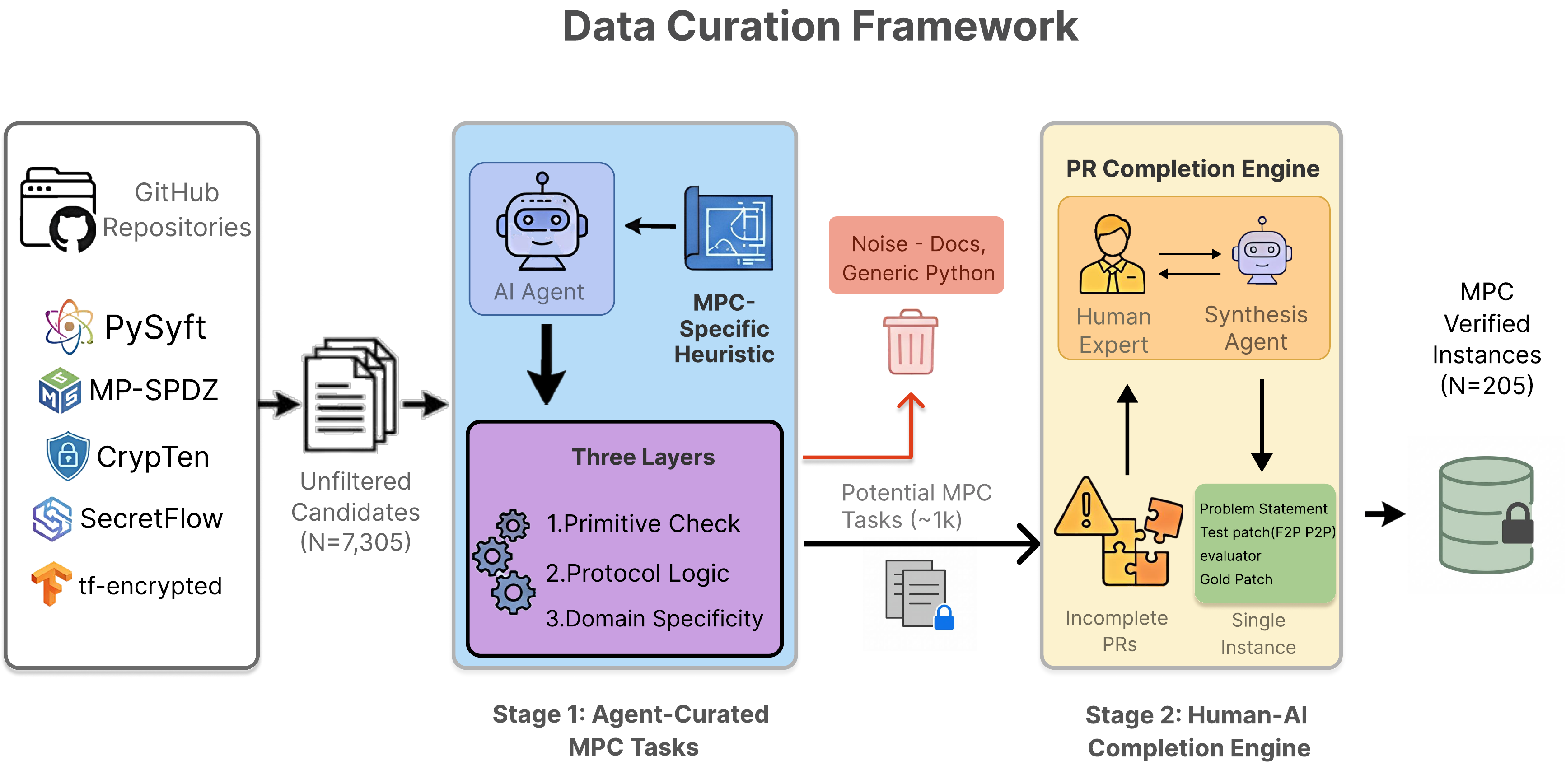}
    \caption{The Data Curation Framework: from 7{,}305 raw GitHub PRs through agent-driven curation and human-AI completion to 205 fully verified MPC task instances.}
    \label{fig:curation_pipeline}
\end{figure*}

\subsection{Data Curation Framework}
\label{sec:data_curation}
We crawl the five MPC frameworks introduced in Section~\ref{sec:bg_mpc} through the GitHub API following the SWE-bench protocol \cite{jimenez2024swebench}, yielding 7{,}305 candidate PRs (Table~\ref{tab:raw-data}). The framework's two stages, depicted in Figure~\ref{fig:curation_pipeline}, are described next.

\subsubsection{Domain-Specific Curation Agent}
\label{sec:curation_agent}
The objective of this stage is to retain, at scale and with principled coverage of the MPC software stack, only those pull requests that exercise genuine MPC reasoning, while emitting a graded confidence so that scarce human review can focus on ambiguous cases rather than obvious noise. Our key observation is that practical MPC software stacks are organised hierarchically---the architectural blueprints of frameworks such as CrypTen \cite{knott2021crypten} expose a bottom-up structure of cryptographic primitives (secret shares, oblivious transfer), interactive protocol logic (e.g., A2B conversion, Beaver-triple synchronisation), and domain-specific secure tensor operations. We therefore align our curator with this hierarchy via three orthogonal stack-aware layers, each interrogating one tier of the stack, and combine their outputs into a graded confidence that drives autonomous routing.

Concretely, an AI Agent processes each candidate's diff and natural-language metadata through a fixed Domain-Specific Heuristic prompt that decomposes into the three layers:

\begin{itemize}
    \item \textbf{Layer 1: Primitive Check.} Verifies direct manipulation of core cryptographic primitives. As highlighted in standard Systematization of Knowledge (SoK) studies \cite{sok_mpc}, practical MPC implementations inherently rely on foundational mathematical constructs at the base layer, such as arithmetic/binary secret sharing and Oblivious Transfer (OT). This layer ensures the code modification directly operates at this cryptographic bedrock.

    \item \textbf{Layer 2: Protocol Logic.} Distinguishes fixes addressing mathematical MPC correctness from generic system wrappers or API formatting. State-of-the-art MPC frameworks heavily rely on the combination of multiple protocols and the secure, efficient conversions between them (e.g., Arithmetic-to-Binary or A2B conversions) \cite{braun2022motion}. This layer evaluates whether the issue tackles the logic and state machines of these multi-party interactive protocols.

    \item \textbf{Layer 3: Domain Specificity.} Ensures the resolution demands specialized cryptographic reasoning rather than general software engineering. This explicitly checks for MPC-exclusive engineering constraints that a standard developer would not encounter, such as maintaining strict data-obliviousness during tensor computations (e.g., within a \texttt{CrypTensor}\cite{knott2021crypten} structure), managing fixed-point truncation errors over finite rings, or optimizing multi-party communication rounds.
\end{itemize}

Each layer maps onto a distinct tier of the MPC stack, so a candidate that activates one or more layers is necessarily exercising MPC-specific code rather than generic Python. The Agent emits a calibrated confidence from the layer outputs, and we apply the autonomous routing of Algorithm~\ref{alg:stage2_curation}: candidates below the noise threshold $\tau_{med}=0.75$ are discarded as generic noise; those above $\tau_{high}=0.9$ are admitted to the potential pool $\mathcal{C}_{potential}$ directly; and the band $[\tau_{med},\tau_{high})$ is routed to a review pool $\mathcal{C}_{review}$ for human adjudication.

\paragraph{Result.} Applied to the 7{,}305 raw candidates, the curator discards 6{,}130 generic-noise PRs and retains 1{,}175 cryptographically relevant tasks (596 high-confidence accepted automatically and 579 medium-confidence routed for human review), which then enter the human-AI completion stage of Section~\ref{sec:human_in_the_loop}.

\subsubsection{Human-AI Completion Engine} \label{sec:human_in_the_loop}

The objective of this stage is to convert the 1{,}175 cryptographically relevant candidates surfaced by curation into standardised, executable benchmark instances, including those whose original pull requests ship without developer-authored Fail-to-Pass (F2P) and Pass-to-Pass (P2P) tests---a population that strict SWE-bench-style filtering would discard, leaving only 42 viable instances out of the same 7{,}305 PRs (Appendix~\ref{app:strict_swe_filtering}). Our observation is that the missing artefacts (clean problem statement, F2P/P2P test patch, executable harness) are tractable to synthesise from the raw PR diff plus expert cryptographic intuition: the diff itself encodes the intended fix, and an expert can identify which behaviours the F2P test must isolate. We therefore introduce a \textbf{Human-AI Collaborative Completion} paradigm in which a human expert and a synthesis agent jointly reconstruct the missing artefacts.

As illustrated in the third stage of Figure~\ref{fig:curation_pipeline}, the 1{,}175 candidates are routed through a PR Completion Engine that operates as a tightly coupled human-AI workflow. Rather than discarding cryptographically sound but incomplete PRs, a human expert co-pilots with a synthesis agent: the expert provides domain-specific intuition and verifies cryptographic correctness, while the agent structures and generates the missing programmatic components. PRs flagged as medium-confidence by curation first pass through a manual offline adjudication step to confirm validity before being admitted to the completion pipeline.

Through this bidirectional collaboration the engine constructs the components of a standard, executable instance---a clear problem statement, the missing F2P/P2P test patch, an evaluator execution harness, and a verified gold patch. The resulting instances treat test-incomplete PRs as completable rather than disqualifying, ensure that synthesised problem statements and tests reflect the actual cryptographic fix (the human expert supplies ground truth), and can be scored against LLM patches automatically without further human intervention.

\paragraph{Result.} The Data Curation Framework distils the 7{,}305 raw candidates into \textbf{205 verified instances}---a 4.9$\times$ expansion over the 42-instance yield of strict SWE-bench filtering on the same corpus (Appendix~\ref{app:strict_swe_filtering}).

\subsection{MPC Verifier}
\label{sec:verification_framework}
\begin{figure}[H]
    \centering
    \includegraphics[width=0.85\textwidth]{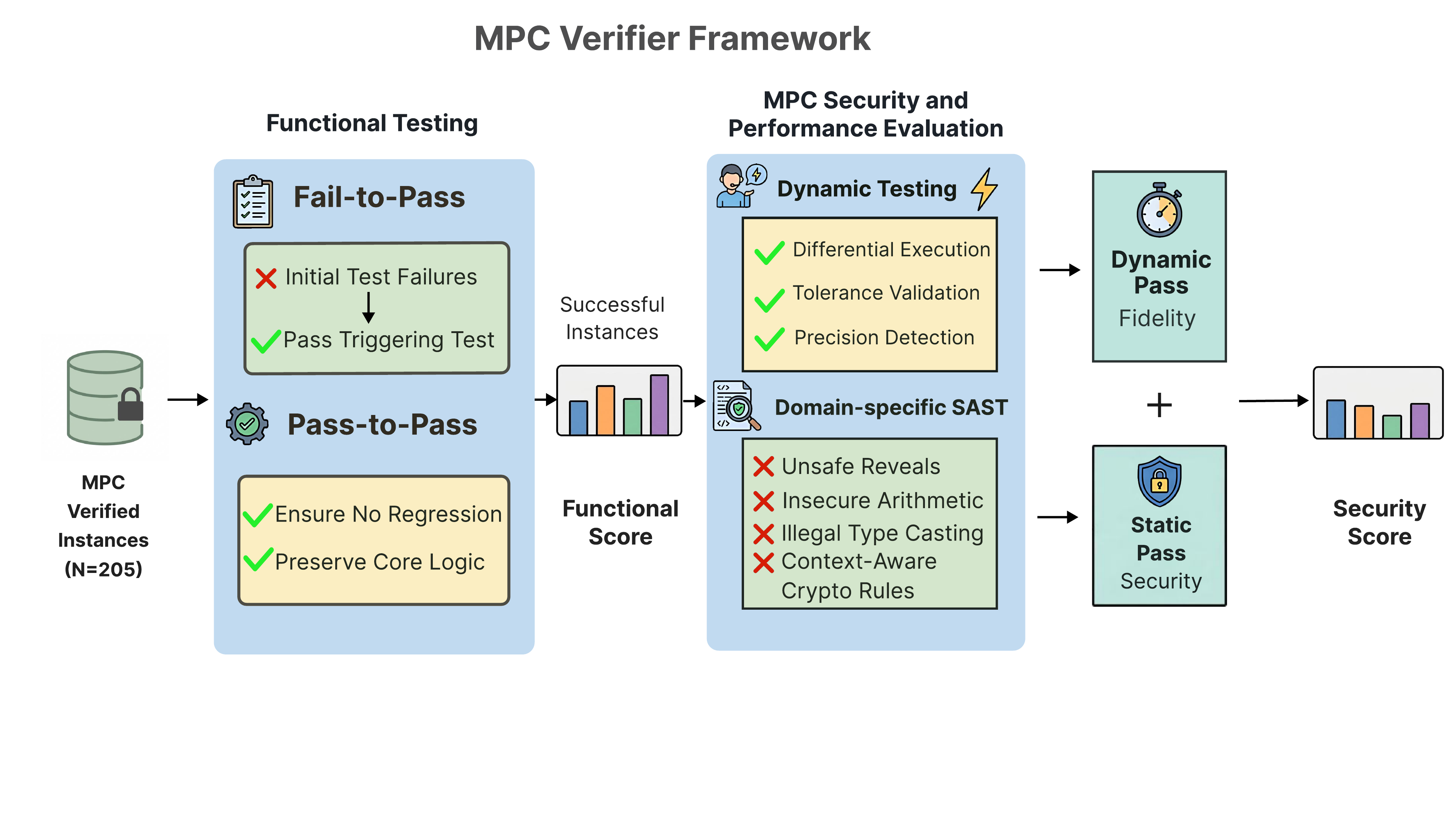}
    \caption{The dual-stream evaluation framework for MPC-Patch-Bench. The framework parallelizes analysis into (1) Dynamic Testing, verifying execution fidelity and numeric precision, and (2) Domain-Specific SAST, performing deep structural security scanning without execution.}
    \label{fig:verifier_pipeline}
\end{figure}

The objective of this stage is to determine whether an LLM-generated patch is not merely functionally correct but also \emph{cryptographically safe} and \emph{numerically faithful}, while remaining practical and reproducible at benchmark scale---a constraint that rules out heavyweight formal proof systems. Our observation is that MPC failure modes separate into two complementary classes: \emph{structural} failures (unsafe reveal patterns, non-cryptographic randomness, illegal public/private casts) that are visible in the source and detectable without execution, and \emph{behavioural} failures (precision drift, edge-case divergences from a plaintext oracle) that only appear when the secure protocol is actually run. We therefore design the verifier as two parallel analytical streams---a \emph{dynamic} stream that runs the protocol against a plaintext oracle and a \emph{static} stream that scans the source for MPC-specific anti-patterns---each producing a score that is combined into a single verified-resolution decision. Figure~\ref{fig:verifier_pipeline} illustrates this dual-stream design.

\subsubsection{Dynamic Testing}
The Dynamic Testing stream orchestrates multi-party code execution and emits a \textbf{Dynamic Score} for execution fidelity through three mechanisms: \emph{differential execution} (via MPCDiff-style comparison of encrypted outputs against a mathematical plaintext oracle), \emph{tolerance validation} (injecting strict boundary values such as $0$, $-1$, $\text{MAX\_INT}$ and complementary inequalities to probe algebraic edge cases), and \emph{precision detection} (monitoring fixed-point truncations and cumulative tensor-operation error against ultra-tight bounds, e.g., $\text{atol} = 0.005$).

\subsubsection{Domain-Specific SAST}
The Domain-Specific Static Application Security Testing (SAST) module performs an exhaustive source-only scan that emits a \textbf{Static Score} for structural security. It flags four classes of MPC anti-pattern: \emph{unsafe reveals} (e.g., \texttt{reveal()} or \texttt{get\_plain\_text()} inside an iterative loop), \emph{insecure arithmetic} (raw floating-point coercion on secret-shared types or non-cryptographic randomness such as \texttt{np.random.rand()}), \emph{illegal type casting} (unauthorised conversions between public and private primitives that silently compromise data-oblivious boundaries), and \emph{context-aware crypto-rule violations} (missing synchronisation steps or breaches of context-dependent security policies).

A worked example of a context-aware SAST rule (Semgrep YAML) and a flagged CrypTen patch is provided in Appendix~\ref{app:sast_example}.

\paragraph{Effect.} The empirical impact of the dual-stream verifier on real LLM-generated patches---reordering the model leaderboard and rejecting up to 40\% of functionally-passing patches as unsafe or numerically degraded---is reported in Section~\ref{sec:verification_impact}.

\section{Dataset}

\subsection{Dataset Overview}

The MPC-Patch-Bench dataset comprises \textbf{205 fully verified and executable task instances}, curated from five foundational open-source MPC libraries: CrypTen \cite{knott2021crypten}, tf-encrypted \cite{dahl2018tfencrypted}, MP-SPDZ \cite{keller2020mp}, SecretFlow \cite{ma2023secretflow}, and PySyft \cite{ziller2021pysyft}. As illustrated in Figure~\ref{fig:dataset_dist}, the dataset spans a diverse range of MPC paradigms: CrypTen (29.8\%, $n=61$) and tf-encrypted (21.0\%, $n=43$) represent tensor-based, ML-oriented secret sharing frameworks; MP-SPDZ (20.0\%, $n=41$) covers protocol-level MPC with multi-party arithmetic; SecretFlow (12.7\%, $n=26$) addresses federated and hybrid MPC computation; and PySyft (16.6\%, $n=34$) represents privacy-preserving deep learning infrastructure. This cross-library diversity ensures that MPC-Patch-Bench evaluates LLMs against a broad and representative spectrum of real-world MPC engineering challenges, rather than overfitting to any single framework or cryptographic paradigm.

\noindent
\begin{minipage}[t]{0.44\textwidth}
    \vspace{0pt}
    \centering
    \includegraphics[width=\linewidth]{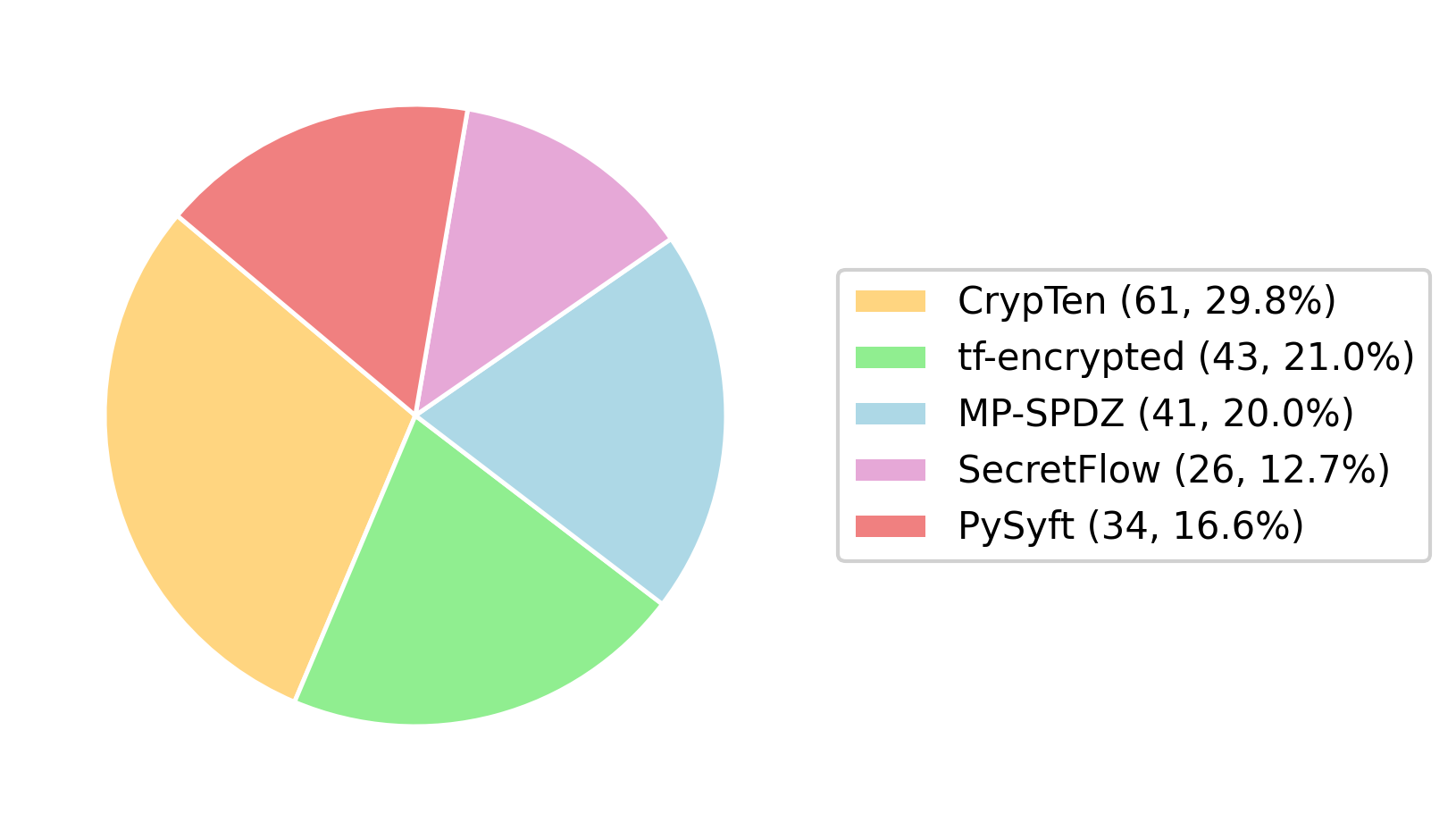}
    \captionof{figure}{Distribution of MPC-Patch-Bench's 205 task instances across five MPC libraries. CrypTen dominates (29.8\%, $n=61$), while SecretFlow contributes the smallest share (12.7\%, $n=26$).}
    \label{fig:dataset_dist}
\end{minipage}
\hfill
\begin{minipage}[t]{0.53\textwidth}
    \vspace{0pt}
    \centering
    \makeatletter\def\@captype{table}\makeatother
    \caption{Statistics of MPC-Patch-Bench task instances (micro-averages across all 205 instances).}
    \label{tab:dataset_stats}
    \small
    \renewcommand{\arraystretch}{1.55}
    \begin{tabular}{llrr}
        \toprule
        & & \textbf{Mean} & \textbf{Max} \\
        \midrule
        \multirow{2}{*}{Problem Statement} & Length (Words) & 47.8  & 481    \\
                                           & Length (Chars) & 372.6 & 3,557  \\
        \midrule
        \multirow{2}{*}{Gold Patch}        & \# Lines edited & 583.5 & 74,717 \\
                                           & \# Files edited &   9.7 &    842 \\
        \midrule
        \multirow{2}{*}{Tests}             & \# F2P tests    &   2.2 &     51 \\
                                           & \# P2P tests    &   1.4 &     64 \\
        \bottomrule
    \end{tabular}
\end{minipage}

\subsection{Complexity and Diversity of Task Characteristics}

\begin{figure}[htbp]
    \centering
    \includegraphics[width=0.7\textwidth]{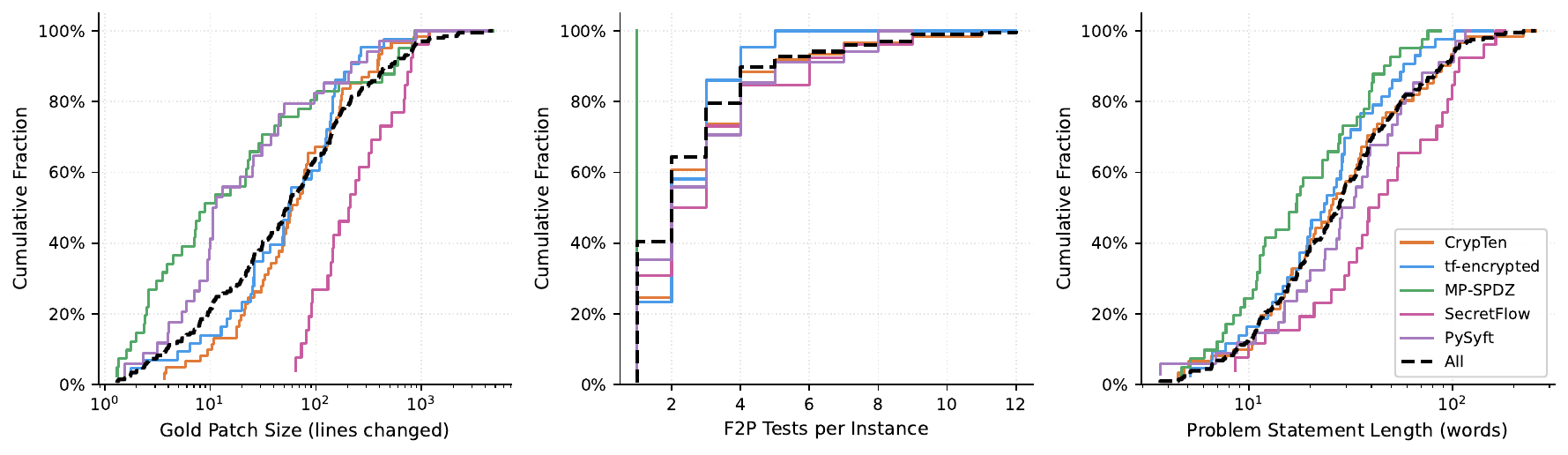}
    \caption{Cumulative distribution functions (CDFs) of MPC-Patch-Bench tasks: (left) gold patch size, (centre) Fail-to-Pass test count per instance, and (right) problem statement word count. See main text for analysis.}
    \label{fig:cdfs}
\end{figure}

To comprehensively evaluate the reasoning span required by our benchmark, we decompose each MPC-Patch-Bench instance into a structured tuple: a \textit{problem statement} (the developer's natural language intent), a \textit{code context} (the repository state), a \textit{test patch} (Fail-to-Pass and Pass-to-Pass verifiers), and a \textit{gold patch} (the empirical solution). Figure~\ref{fig:cdfs} visualizes the empirical cumulative distribution functions (CDFs) across three critical structural dimensions, demonstrating that MPC-Patch-Bench exhibits both the structural diversity of real-world software engineering and the unique, exacting complexity of cryptographic development.

The CDFs reveal three things at once. \textbf{Gold patch size} (left) ranges from surgical 10--50-line edits typical of MP-SPDZ to long tails beyond 200 lines (prominent in SecretFlow), confirming that MPC-Patch-Bench captures the full spectrum of cryptographic-software complexity rather than isolated functions. \textbf{Fail-to-Pass test count} (centre) clusters tightly across all five frameworks---more than half of all instances (55\%) are gated by a single rigorously constrained F2P test, exposing a domain-universal hallmark: MPC correctness is verified against precise mathematical invariants rather than expansive behavioural suites, imposing a steep penalty on iterative guessing. \textbf{Problem statement length} (right) varies widely (under 10 words to over 100), spanning terse bug reports requiring deep repository-level context to extensive cryptographic discussions demanding step-by-step reasoning, preventing models from overfitting to a specific prompt formulation.

A complementary analysis of how task difficulty---measured by per-instance resolution rate---varies with gold-patch size and Fail-to-Pass test count is presented in Appendix~\ref{app:difficulty}.

\section{Evaluation}

\subsection{Experimental Setup}
\label{sec:experimental_setup}

\noindent\textbf{Models and prompting.} We evaluate eight contemporary LLMs spanning Anthropic, OpenAI, and Google (Sonnet 4.6, Opus 4.6, Haiku 4.5, GPT-5.4, GPT-4.1, GPT-4o-mini, Gemini 2.5 Pro, Gemini 2.5 Flash-Lite; see Appendix~\ref{app:reproducibility} for exact API IDs). All models are queried zero-shot at temperature 0 where supported, with a 16{,}384-token output cap (65{,}536 for Gemini); requests exceeding the model's context window are logged and excluded from resolution statistics. Each prompt provides the MPC-related problem statement, the relevant pre-patch source file(s), and instructions to emit complete modified files using a \texttt{<<<FILE: path>>>...<<<END>>>} delimiter format.

\noindent\textbf{Test execution.} Generated patches are applied to the repository at the base commit and executed using library-specific test runners---\texttt{pytest} with a 300-second per-test timeout for CrypTen, tf-encrypted, SecretFlow, and PySyft, and a custom \texttt{compile.py}~+~\texttt{emulate.sh} pipeline for MP-SPDZ. An instance is marked \emph{functionally resolved} iff all F2P tests pass and all P2P tests continue to pass; resolved patches additionally undergo the dual-stream verifier of Section~\ref{sec:verification_framework} for the verified-resolution metric. All runs use an HPC cluster (RHEL~8) with per-instance isolation via separate git worktrees.

\subsection{Functional Resolution Rates}
\label{sec:functional_resolution}
\begin{figure}[htbp]
    \centering
    \includegraphics[width=0.7\textwidth]{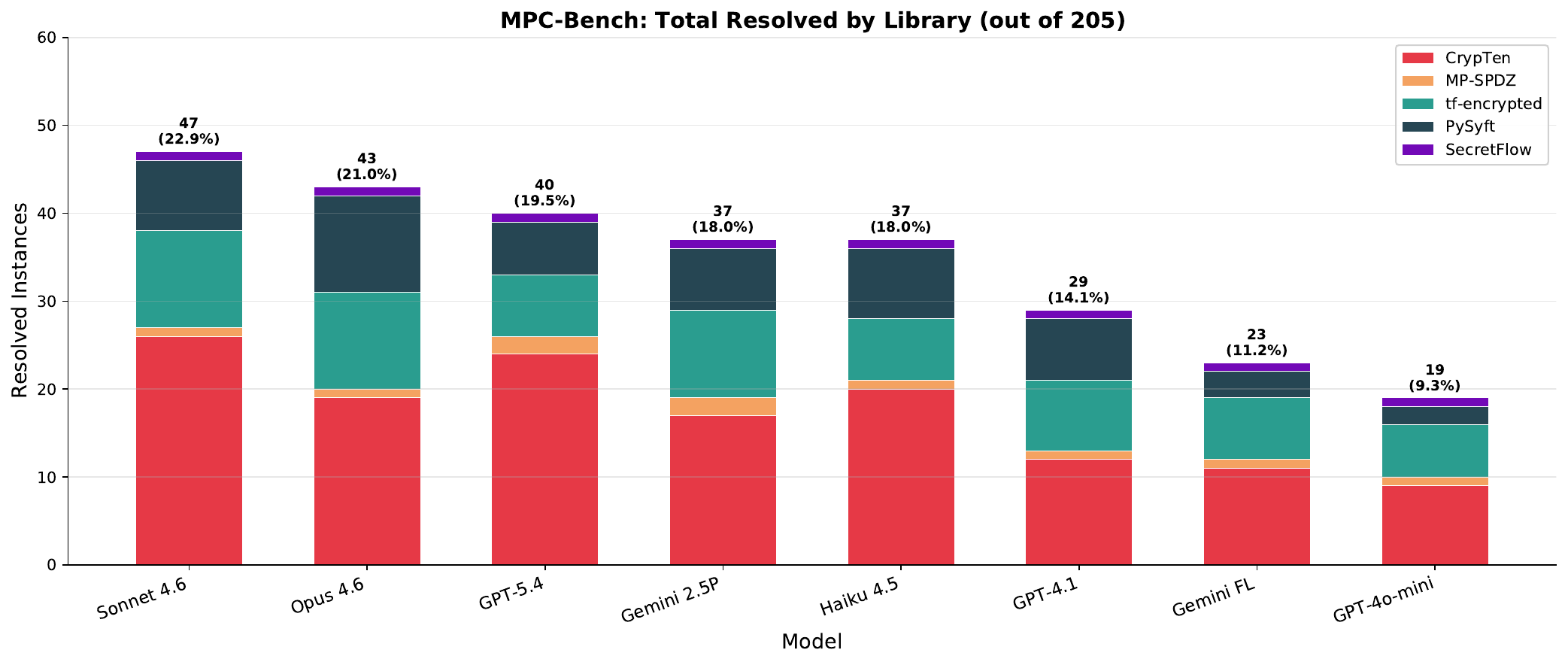}
    \caption{MPC-Patch-Bench model ranking and overall functional resolution rates across 205 task instances. We evaluate eight leading models: Sonnet 4.6, Opus 4.6, GPT-5.4, Gemini 2.5P, Haiku 4.5, GPT-4.1, Gemini FL, and GPT-4o-mini. The results reveal inter-library variation in difficulty, with Sonnet 4.6 achieving the highest overall functional resolution rate of 22.9\%.}
    \label{fig:model_ranking}
\end{figure}

As Figure~\ref{fig:model_ranking} shows, even the strongest evaluated model resolves only 22.9\% of the 205 tasks, and difficulty varies sharply across the five MPC frameworks---high-level PyTorch-style libraries are comparatively tractable while protocol-level systems remain almost entirely unresolved by every evaluated model. We unpack this aggregate picture by model and by library below.

Anthropic's Claude family leads the ranking---notably, the lighter Sonnet 4.6 (22.9\%) outperforms the heavier Tier-1 models Opus 4.6 (21.0\%) and GPT-5.4 (19.5\%). The remaining models cluster from mid-teens (Gemini 2.5P / Haiku 4.5 at 18.0\%, GPT-4.1 at 14.1\%) down to $\sim$10\% for the smallest (Gemini FL, GPT-4o-mini). The per-library breakdown is sharply bimodal: CrypTen is consistently the most tractable framework, likely because its PyTorch-aligned APIs overlap closely with LLM pretraining data, while MP-SPDZ at the opposite extreme remains nearly unsolved (top models resolve at most 2 of its 41 instances), reflecting the difficulty of its bespoke, C++-like scripting language and low-level arithmetic protocol logic. SecretFlow, tf-encrypted, and PySyft fall between.

Collectively, these results confirm that MPC-Patch-Bench constitutes a rigorous and domain-discriminative benchmark: even the best-performing models on general software engineering tasks struggle substantially when confronted with authentic MPC cryptographic engineering challenges.

\subsection{Verification Impact on Model Performance}
\label{sec:verification_impact}
\begin{figure}[htbp]
    \centering
    \includegraphics[width=0.7\textwidth]{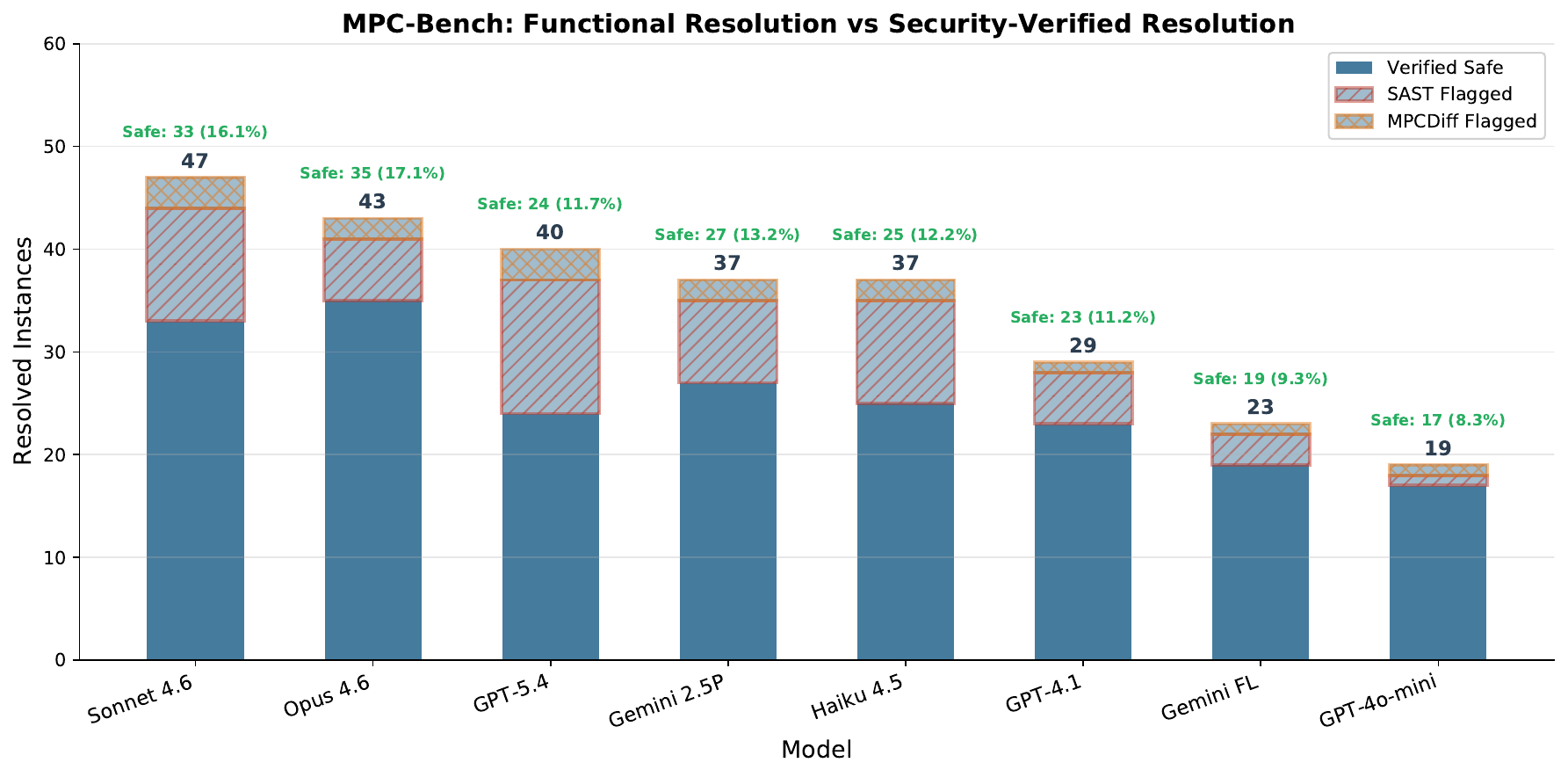}
    \caption{Functional vs verified resolution per model on MPC-Patch-Bench, with attrition split into SAST and MPCDiff (Dynamic) flags.}
    \label{fig:security_filter}
\end{figure}

As Figure~\ref{fig:security_filter} shows, applying the dual-stream verifier of Section~\ref{sec:verification_framework} to every functionally-resolved patch substantively reorders the leaderboard and rejects between 11\% and 40\% of functionally-passing patches across models, demonstrating that functional resolution and verified resolution are materially different signals in the MPC setting.

Sonnet 4.6 loses 14 of its 47 functional resolutions (29\% attrition, 11 SAST + 3 MPCDiff flags) and drops to 16.1\% verified, while Opus 4.6 loses only 8 of its 43 (retaining 35, 17.1\% verified) and overtakes Sonnet on the verified leaderboard; GPT-5.4 sees the most severe drop, losing 16 of its 40 (40\% rejection, 13 SAST + 3 MPCDiff flags). The reordering confirms that functional ranking and verified ranking are materially different signals: a non-trivial fraction of patches passing standard tests still fail security or numerical-fidelity checks, and the gap is large enough to invert the model leaderboard.

\section{Conclusion}
\label{sec:conclusion}

We presented MPC-Patch-Bench, the first repository-level benchmark for evaluating LLM code repair on real-world MPC software. By coupling a Data Curation Framework with an MPC Verifier, MPC-Patch-Bench surfaces capability gaps that general-purpose benchmarks miss: functional correctness alone substantially overestimates LLM reliability in privacy-preserving software, since a non-trivial fraction of test-passing patches still fail security or numerical-fidelity checks. We hope MPC-Patch-Bench serves as a methodological template for benchmarking LLMs in domains where correctness is determined by more than test outcomes.

\paragraph{Limitations and future work.} The current release covers five widely-used MPC frameworks; extending the benchmark to additional cryptographic stacks (e.g., emerging fully-homomorphic-encryption hybrids and zero-knowledge programming environments) is a natural next step.

\newpage
\bibliographystyle{alpha}
\bibliography{sample}

@inproceedings{jimenez2024swebench,
  title     = {{SWE}-bench: Can Language Models Resolve Real-world Github Issues?},
  author    = {Carlos E Jimenez and John Yang and Alexander Wettig and Shunyu Yao and Kexin Pei and Ofir Press and Karthik R Narasimhan},
  booktitle = {The Twelfth International Conference on Learning Representations},
  year      = {2024},
  url       = {https://openreview.net/forum?id=VTF8yNQM66}
}

@inproceedings{yang2025swesmith,
  title     = {{SWE}-smith: Scaling Data for Software Engineering Agents},
  author    = {Yang, John and Lieret, Kilian and Jimenez, Carlos E. and Wettig, Alexander and Khandpur, Kabir and Zhang, Yanzhe and Hui, Binyuan and Press, Ofir and Schmidt, Ludwig and Yang, Diyi},
  booktitle = {Advances in Neural Information Processing Systems (NeurIPS)},
  year      = {2025},
  url       = {https://arxiv.org/abs/2504.21798}
}

@inproceedings{pang2024mpcdiff,
  title     = {{MPCDiff}: Testing and Repairing {MPC}-Hardened Deep Learning Models},
  author    = {Pang, Qi and Yuan, Yuanyuan and Wang, Shuai},
  booktitle = {31st Annual Network and Distributed System Security Symposium (NDSS)},
  year      = {2024},
  url       = {https://www.ndss-symposium.org/ndss-paper/mpcdiff-testing-and-repairing-mpc-hardened-deep-learning-models/}
}

@article{dong2024spdzcoder,
  title   = {{SPDZCoder}: Combining Expert Knowledge with {LLMs} for Generating Privacy-Computing Code},
  author  = {Dong, Xiaoning and Xin, Peilin and Xu, Wei},
  journal = {arXiv preprint arXiv:2501.00363},
  year    = {2024},
  url     = {https://arxiv.org/abs/2501.00363}
}

@misc{chen2021evaluatinglargelanguagemodels,
  title         = {Evaluating Large Language Models Trained on Code},
  author        = {Mark Chen and Jerry Tworek and Heewoo Jun and Qiming Yuan and Henrique Ponde de Oliveira Pinto and Jared Kaplan and Harri Edwards and Yuri Burda and Nicholas Joseph and Greg Brockman and Alex Ray and Raul Puri and Gretchen Krueger and Michael Petrov and Heidy Khlaaf and Girish Sastry and Pamela Mishkin and Brooke Chan and Scott Gray and Nick Ryder and Mikhail Pavlov and Alethea Power and Lukasz Kaiser and Mohammad Bavarian and Clemens Winter and Philippe Tillet and Felipe Petroski Such and Dave Cummings and Matthias Plappert and Fotios Chantzis and Elizabeth Barnes and Ariel Herbert-Voss and William Hebgen Guss and Alex Nichol and Alex Paino and Nikolas Tezak and Jie Tang and Igor Babuschkin and Suchir Balaji and Shantanu Jain and William Saunders and Christopher Hesse and Andrew N. Carr and Jan Leike and Josh Achiam and Vedant Misra and Evan Morikawa and Alec Radford and Matthew Knight and Miles Brundage and Mira Murati and Katie Mayer and Peter Welinder and Bob McGrew and Dario Amodei and Sam McCandlish and Ilya Sutskever and Wojciech Zaremba},
  year          = {2021},
  eprint        = {2107.03374},
  archiveprefix = {arXiv},
  primaryclass  = {cs.LG},
  url           = {https://arxiv.org/abs/2107.03374}
}

@inproceedings{austin2021mbpp,
  title   = {Program Synthesis with Large Language Models},
  author  = {Austin, Jacob and Odena, Augustus and Nye, Maxwell and Bosma, Maarten and Michalewski, Henryk and others},
  journal = {arXiv preprint arXiv:2108.07732},
  year    = {2021}
}

@inproceedings{liu2023evalplus,
  title     = {Is Your Code Generated by {ChatGPT} Really Correct? Rigorous Evaluation of Large Language Models for Code Generation},
  author    = {Liu, Jiawei and Xia, Chunqiu Steven and Wang, Yuyao and Zhang, Lingming},
  booktitle = {Thirty-seventh Conference on Neural Information Processing Systems (NeurIPS)},
  year      = {2023}
}

@article{yu2024codereval,
  title   = {{CoderEval}: A Benchmark of Pragmatic Code Generation with Generative Pre-trained Models},
  author  = {Yu, Hao and Shen, Bo and Ran, Dezhi and Zhang, Jiaxin and Zhang, Qiang and others},
  journal = {ICSE 2024},
  year    = {2024}
}

@inproceedings{lai2023ds1000,
  title     = {{DS-1000}: A Natural and Reliable Benchmark for Data Science Code Generation},
  author    = {Lai, Yuhang and Li, Chengxi and Wang, Yiming and Zhang, Tianyi and Zhong, Ruiqi and Zettlemoyer, Luke and Yih, Wen-tau and Fried, Daniel and Wang, Sida and Yu, Tao},
  booktitle = {International Conference on Machine Learning (ICML)},
  year      = {2023}
}

@article{liu2025cryptorch,
  title   = {{CrypTorch}: {PyTorch}-based Auto-tuning Compiler for Machine Learning with Multi-party Computation},
  author  = {Liu, Jinyu and Tan, Gang and Maeng, Kiwan},
  journal = {arXiv preprint arXiv:2511.19711},
  year    = {2025},
  url     = {https://arxiv.org/abs/2511.19711}
}

@inproceedings{knott2021crypten,
  title     = {{CrypTen}: Secure Multi-Party Computation Meets Machine Learning},
  author    = {Knott, Brian and Venkataraman, Shobha and Hannun, Awni and Sengupta, Shubho and Ibrahim, Mark and van der Maaten, Laurens},
  booktitle = {Advances in Neural Information Processing Systems (NeurIPS)},
  year      = {2021},
  url       = {https://papers.neurips.cc/paper/2021/hash/2754518221cfbc8d25c13a06a4cb8421-Abstract.html}
}

@article{ziller2021pysyft,
  title   = {{PySyft}: A Library for Easy Federated Learning},
  author  = {Ziller, Alexander and Trask, Andrew and Lopardo, Antonio and others},
  journal = {arXiv preprint arXiv:2105.05134},
  year    = {2021},
  url     = {https://arxiv.org/abs/2105.05134}
}

@inproceedings{keller2020mp,
  title     = {{MP-SPDZ}: A Versatile Framework for Multi-Party Computation},
  author    = {Keller, Marcel},
  booktitle = {Proceedings of the 2020 ACM SIGSAC Conference on Computer and Communications Security (CCS)},
  pages     = {1575--1590},
  year      = {2020},
  url       = {https://eprint.iacr.org/2020/521}
}

@article{li2024metamorphic,
  title     = {Metamorphic Testing of Secure Multi-Party Computation (MPC) Compilers},
  author    = {Li, Yichen and Xiao, Dongwei and Liu, Zhibo and Pang, Qi and Wang, Shuai},
  journal   = {Proceedings of the ACM on Software Engineering},
  volume    = {1},
  number    = {FSE},
  pages     = {1216--1237},
  year      = {2024},
  publisher = {ACM New York, NY, USA},
  doi       = {10.1145/3639477.3640030},
  url       = {https://dl.acm.org/doi/10.1145/3639477.3640030}
}

@inproceedings{ruan2025hawkeye,
  title     = {{HawkEye}: Statically and Accurately Profiling the Communication Cost of Models in Multi-party Learning},
  author    = {Ruan, Wenqiang and Lin, Xin and Zhou, Ruisheng and Lin, Guopeng and Yu, Shui},
  booktitle = {34th USENIX Security Symposium (USENIX Security)},
  year      = {2025},
  url       = {https://www.usenix.org/conference/usenixsecurity25/presentation/ruan}
}

@inproceedings{ma2023secretflow,
  title     = {{SecretFlow-SPU}: A Performant and User-Friendly Framework for Privacy-Preserving Machine Learning},
  author    = {Ma, Junming and Zheng, Yancheng and Feng, Jun and Zhao, Derun and Wu, Haoqi and Fang, Wenjing and Tan, Jin and Yu, Chaofan and Zhang, Benyu and Wang, Lei},
  booktitle = {2023 USENIX Annual Technical Conference (USENIX ATC)},
  year      = {2023},
  url       = {https://www.usenix.org/conference/atc23/presentation/ma}
}

@misc{dahl2018tfencrypted,
  title         = {Private Machine Learning in {TensorFlow} using Secure Computation},
  author        = {Dahl, Morten and Mancuso, Jason and Dupis, Yann and Decoste, Ben and Giraud, Morgan and Livingstone, Ian and Patriquin, Justin and Uhma, Gavin},
  year          = {2018},
  eprint        = {1810.08130},
  archivePrefix = {arXiv},
  primaryClass  = {cs.CR},
  url           = {https://arxiv.org/abs/1810.08130},
  note          = {Privacy Preserving Machine Learning Workshop at NeurIPS 2018}
}

@inproceedings{sok_mpc,
  title        = {SoK: General purpose compilers for secure multi-party computation},
  author       = {Hastings, Marcella and Evans, David and Katz, Jonathan and Wies, Thomas},
  booktitle    = {2019 IEEE Symposium on Security and Privacy (SP)},
  pages        = {1220--1237},
  year         = {2019},
  organization = {IEEE}
}

@article{braun2022motion,
  title={MOTION--a framework for mixed-protocol multi-party computation},
  author={Braun, Lennart and Demmler, Daniel and Schneider, Thomas and Tkachenko, Oleksandr},
  journal={ACM Transactions on Privacy and Security (TOPS)},
  volume={25},
  number={2},
  pages={1--35},
  year={2022},
  publisher={ACM New York, NY}
}

@inproceedings{mohassel2017secureml,
  title={SecureML: A System for Scalable Privacy-Preserving Machine Learning},
  author={Mohassel, Payman and Zhang, Yupeng},
  booktitle={2017 IEEE Symposium on Security and Privacy (SP)},
  pages={19--38},
  year={2017},
  organization={IEEE}
}

@article{wagh2019securenn,
  title={SecureNN: 3-Party Secure Computation for Neural Network Training},
  author={Wagh, Sameer and Gupta, Divya and Chandran, Nishanth},
  journal={Proceedings on Privacy Enhancing Technologies},
  volume={2019},
  number={3},
  pages={26--49},
  year={2019},
  publisher={De Gruyter Open}
}

@inproceedings{zheng2019helen,
  title={Helen: Maliciously Secure Coopetitive Learning for Linear Models},
  author={Zheng, Wenting and Popa, Raluca Ada and Gonzalez, Joseph E and Stoica, Ion},
  booktitle={2019 IEEE Symposium on Security and Privacy (SP)},
  pages={724--738},
  year={2019},
  organization={IEEE}
}

@article{hie2018realizing,
  title={Realizing private and practical pharmacological collaboration},
  author={Hie, Brian and Cho, Hyunghoon and Berger, Bonnie},
  journal={Science},
  volume={362},
  number={6412},
  pages={347--350},
  year={2018},
  publisher={American Association for the Advancement of Science}
}

@inproceedings{bogdanov2016students,
  title={Students and Taxes: a Privacy-Preserving Study Using Secure Computation},
  author={Bogdanov, Dan and Kamm, Liina and Kubo, Baldur and Rebane, Reimo and Vaht, Ville and Willemson, Jan},
  booktitle={Proceedings on Privacy Enhancing Technologies},
  volume={2016},
  number={3},
  pages={117--135},
  year={2016}
}

@inproceedings{rastogi2014wysteria,
  title={Wysteria: A Programming Language for Generic, Mixed-Mode Multiparty Computations},
  author={Rastogi, Aseem and Hammer, Matthew A and Hicks, Michael},
  booktitle={2014 IEEE Symposium on Security and Privacy (SP)},
  pages={655--670},
  year={2014},
  organization={IEEE}
}

@misc{zahur2015oblivc,
  title={Obliv-{C}: A Language for Extensible Data-Oblivious Computation},
  author={Zahur, Samee and Evans, David},
  howpublished={Cryptology ePrint Archive, Report 2015/1153},
  year={2015},
  url={https://eprint.iacr.org/2015/1153}
}

\appendix
\section{Impact of Strict SWE-bench Filtering on MPC Repositories} \label{app:strict_swe_filtering}

To empirically justify the necessity of our Human-AI Collaborative Synthesis (Section \ref{sec:human_in_the_loop}), we conducted a preliminary experiment applying the strict, unmodified SWE-bench data curation pipeline to our collected corpus of 7,305 MPC Pull Requests. 

The standard SWE-bench methodology strictly requires PRs to inherently contain both the code modification and a perfectly parsable, execution-ready test suite (specifically Fail-to-Pass tests) to automatically verify correctness. When this rigid heuristic was applied to our raw dataset, the candidate pool plummeted from 7,305 to only \textbf{42} viable instances. 

Detailed bottleneck analysis across our target repositories reveals that this data loss stems from a systemic mismatch between standard open-source development paradigms and the specific maintenance habits of the MPC community. The attrition is primarily driven by three factors:

\begin{itemize}
    \item \textbf{Non-Standard and Internal Workflows:} Repositories backed by corporate entities often utilize opaque internal tracking systems. For instance, \textit{CrypTen} relies on Meta's internal workflow, resulting in zero merged PRs that can be identified and utilized by standard automated scrapers.
    \item \textbf{Atypical Issue-PR Linking Practices:} Many MPC maintainers favor direct bug pushes over formalized issue-PR workflows. \textit{emp-tool} frequently bypasses standard issue-PR flows entirely, while \textit{MP-SPDZ} demonstrates a high merge rate but negligible issue referencing (less than 5\%). Even \textit{PySyft}, our largest data source, is severely bottlenecked by a mere 6.2\% link rate.
    \item \textbf{Critical Lack of Execution-Ready Tests:} Even when issues are properly linked to PRs, the linked PRs frequently lack the strict test patches required by the SWE-bench format. \textit{SecretFlow} is one such case: a high volume of merged PRs but a comparatively small fraction shipping with associated developer-authored test patches.
\end{itemize}

Consequently, relying solely on the legacy SWE-bench pipeline yields an unrepresentatively small and heavily biased benchmark. This empirical reality fundamentally necessitates our Human-AI Collaborative Synthesis approach to rescue, standardize, and augment these high-value cryptographic challenges.
\begin{table}[htbp]
\centering
\caption{Statistics for how many candidate task instances were kept after the completion of a stage across the construction and validation procedures.}
\label{tab:swe-bench-filtering}
\small
\begin{tabular}{lrrrr}
\toprule
\textbf{Repo} & \textbf{Total PRs Crawled} & \textbf{Post-Conversion} & \textbf{Post-Curation} & \textbf{Post-Validation (Final)} \\
\midrule
crypten & 236 & 0 & 0 & 0 \\
secretflow & 658 & 22& 2 & 0 \\
MP-SPDZ & 110 & 4 & 1 & 1 \\
PySyft & 5,840 & 277 & 159 & 30 \\
tf-encrypted& 461 & 40 & 22 & 11 \\
\midrule
\textbf{Total} & \textbf{7,305} & \textbf{343} & \textbf{184} & \textbf{42} \\
\bottomrule
\end{tabular}
\end{table}

\begin{description}
    \item[Total PRs Crawled] The initial dataset containing all raw, merged pull requests (PRs) harvested from the target GitHub repositories via the developer API.
    
    \item[Post-Conversion] The first filtering stage where PRs are retained only if they are explicitly linked to a GitHub issue and include modifications to test files, ensuring a baseline for verifiability.
    
    \item[Post-Curation] A domain-specific refinement stage in section \ref{sec:human_in_the_loop} where general infrastructure-related tasks (such as CI/CD pipelines, environment setup, or documentation) are filtered out to ensure the remaining instances are strictly focused on core Multi-Party Computation (MPC) logic.
    
    \item[Post-Validation (Final)] The finalized benchmark suite consisting of tasks that passed the execution harness. This confirms that the repository can be successfully installed and that the provided patch resolves a verified "Fail-to-Pass" test case.
\end{description}

\section{Repository Selection and License Compliance} \label{app:compliance}

In addition to the semantic filtering criteria discussed in Section 2, we strictly ensure that all selected MPC repositories are governed by licenses that permit non-proprietary use, academic research, code modification, and redistribution. 

The majority of the software licenses in our curated corpus are highly permissive (e.g., Apache License 2.0, MIT License). For repositories operating under custom or academic-specific licenses, we manually inspected the license terms to confirm they explicitly authorize the benchmarking and research use cases exercised in our work. The specific licenses associated with each core repository evaluated in MPC-Patch-Bench are fully detailed in Table \ref{tab:licenses}.

\begin{table}[htbp]
    \centering
    \renewcommand{\arraystretch}{1.5}
    \begin{tabular}{p{0.3\linewidth} p{0.65\linewidth}}
        \toprule
        \textbf{License} & \textbf{Repository / Description} \\
        \midrule
        \textbf{Apache License 2.0} & \texttt{OpenMined/PySyft}, \texttt{secretflow/secretflow}, \texttt{tf-encrypted/tf-encrypted} \\
        \textbf{MIT License} & \texttt{facebookresearch/CrypTen} \\
        \textbf{Open-Source / Academic License} & \texttt{data61/MP-SPDZ} \newline \textit{Note: Publicly available under an academic license that explicitly permits protocol benchmarking and research extensions.} \\
        \bottomrule
    \end{tabular}
    \caption{Licenses associated with each repository included in MPC-Patch-Bench. All licenses are permissive and allow for public, academic, and non-profit research use.}
    \label{tab:licenses}
\end{table}

\section{Reproducibility} \label{app:reproducibility}

To ensure full reproducibility of our evaluation results, we report the exact model identifiers, API configurations, and evaluation protocol used.

\paragraph{Model Identifiers and API Details.}
All evaluations were conducted between March and April 2026 via official vendor APIs. Table~\ref{tab:model_ids} lists the exact model IDs queried.

\begin{table}[htbp]
\centering
\small
\begin{tabular}{llll}
\toprule
\textbf{Model (Display Name)} & \textbf{API Model ID} & \textbf{Vendor} & \textbf{API Date} \\
\midrule
Sonnet 4.6 & \texttt{claude-sonnet-4-6} & Anthropic & Mar 2026 \\
Opus 4.6 & \texttt{claude-opus-4-6} & Anthropic & Mar 2026 \\
Haiku 4.5 & \texttt{claude-haiku-4-5-20251001} & Anthropic & Mar 2026 \\
GPT-5.4 & \texttt{gpt-5.4} & OpenAI & Mar 2026 \\
GPT-4.1 & \texttt{gpt-4.1} & OpenAI & Mar 2026 \\
GPT-4o-mini & \texttt{gpt-4o-mini} & OpenAI & Mar 2026 \\
Gemini 2.5 Pro & \texttt{gemini-2.5-pro} & Google & Apr 2026 \\
Gemini FL & \texttt{gemini-2.5-flash-lite} & Google & Apr 2026 \\
\bottomrule
\end{tabular}
\caption{Exact model identifiers used in all MPC-Patch-Bench evaluations.}
\label{tab:model_ids}
\end{table}

\paragraph{Generation Parameters.}
All models were queried with a temperature of $0$ (greedy decoding) where supported. For OpenAI's GPT-5.x series, temperature is not configurable via the API and defaults to the model's internal setting. Maximum output length was set to 16{,}384 tokens (Anthropic and OpenAI) and 65{,}536 tokens (Gemini). No system prompts beyond the task-specific evaluation prompt were used. For OpenAI models encountering rate limits (HTTP 429), exponential backoff was applied (initial wait 30s, max 300s). Requests exceeding the model's context window were logged as \texttt{request\_too\_large} and excluded from resolution statistics.

\paragraph{Evaluation Prompt Template.}
Each model received an identical structured prompt consisting of:
\begin{enumerate}
    \item A MPC based problem statement (derived from the linked GitHub issue).
    \item The relevant source file(s) from the repository at the pre-patch commit.
    \item Instructions to output the complete modified file(s) using the delimiter format: \texttt{<<<FILE: path>>>...<<<END>>>}.
\end{enumerate}
No few-shot examples or chain-of-thought elicitation were used; all evaluations are strictly zero-shot.

\paragraph{Execution and Validation Protocol.}
For each generated patch:
\begin{enumerate}
    \item The patch is applied to the repository checked out at the base commit.
    \item The Fail-to-Pass (F2P) and Pass-to-Pass (P2P) test suites are executed using the library-specific test runner:
    \begin{itemize}
        \item \textbf{CrypTen, tf-encrypted, SecretFlow, PySyft:} \texttt{pytest} with a 300-second timeout per test.
        \item \textbf{MP-SPDZ:} Custom compile-and-emulate pipeline (\texttt{python3 compile.py <test> \&\& Scripts/emulate.sh <test>}).
    \end{itemize}
    \item An instance is marked \textit{resolved} if and only if all F2P tests pass and all P2P tests continue to pass.
    \item For the security-aware evaluation (Section~\ref{sec:verification_impact}), resolved patches additionally undergo the dual-stream SAST + Dynamic Testing verification.
\end{enumerate}

\paragraph{Computational Resources.}
All test execution was performed on an HPC cluster running RHEL~8 (Linux 4.18) with per-instance isolation via separate \texttt{git worktree}s. CrypTen's secure-tensor runtime requires a CUDA-capable GPU, for which we used the cluster's NVIDIA V100~32GB / H100~PCIe / H100~80GB~HBM3 partitions; the other libraries (tf-encrypted, SecretFlow, PySyft, MP-SPDZ) ran CPU-only. Cumulative vendor-API spend across the eight evaluated LLMs on the 205-instance benchmark was on the order of a few hundred US dollars.

\section{Data Contamination Analysis} \label{app:contamination}

A concern for benchmarks derived from public repositories is potential memorization of patches during pretraining. To assess the potential impact of data contamination on MPC-Patch-Bench, we examine the benchmark along three orthogonal dimensions.

\paragraph{Task Characteristics Limit the Value of Memorisation.}
Each instance presents the model with a buggy codebase snapshot at a specific commit and a natural-language bug report; the model must \textit{comprehend the fault} and produce a correct fix. Even if a model has memorized the original PR diff during pretraining, this knowledge alone is insufficient: the fix must be consistent with the exact repository state at the base commit, and the bug report is often rephrased or synthesized by PR completion engine, so rote recall of a memorized patch cannot substitute for genuine fault localization and repair. MPC tasks further require exact finite-field arithmetic and protocol-compliant communication—a single wrong constant causes test failure, leaving no room for approximate recall. Moreover, 163 of 205 instances were constructed via our Human-AI Synthesis pipeline with custom problem statements and test patches that do not exist verbatim in any public repository.

\paragraph{Headline Resolution Rates Are Far Below What Pure Recall Would Predict.}
A model that simply reproduced the upstream gold patch verbatim from pretraining would resolve every instance whose patch it had memorised. Instead, the strongest evaluated model resolves only 22.9\% of instances, and resolution drops to 17.1\% under the MPC Verifier; the absolute resolution ceiling is therefore inconsistent with patch recall as the dominant mechanism. Moreover, 11\%--40\% of functionally-passing patches fail security verification: models relying on memorised gold patches would pass cleanly, so the substantial verifier attrition rate suggests models generate novel (and sometimes insecure) solutions rather than retrieving stored ones.

\paragraph{A Reusable Methodology for Future Benchmarks.}
Our curation pipeline---the three-layer curation agent, the PR completion engine, and the oracle validation framework---is entirely repository-agnostic: it requires only a git repository with commit history and makes no assumptions about public availability. We therefore view it as a reusable methodology rather than a one-off artifact: teams with access to internal or proprietary MPC codebases can apply the same procedure to construct benchmarks aligned with their own code and threat models, and the wider community can re-run it on newer commit ranges as repositories evolve. We do not claim that any single benchmark, MPC-Patch-Bench included, can fully escape contamination; the practical value of releasing the methodology is that it lowers the cost of producing fresh evaluation instances whenever existing ones become saturated or suspect, keeping contamination effects bounded over time without requiring access to wholly private data.

\section{Domain-Specific Curation Agent: Algorithm Pseudocode} \label{app:curation_alg}

The autonomous routing logic of the Domain-Specific Curation Agent (Section~\ref{sec:curation_agent}) is formalised in Algorithm~\ref{alg:stage2_curation}.

\begin{algorithm}[H]
\caption{Domain-Specific Curation Agent}
\label{alg:stage2_curation}
\begin{algorithmic}[1]
\REQUIRE Unfiltered Candidates $\mathcal{C}_{raw}$, AI Agent $\mathcal{A}_{agent}$, Fixed MPC-Specific Heuristic $\mathcal{H}_{mpc}$, Confidence thresholds $\tau_{high}$ and $\tau_{med}$ (currently we set $\tau_{high} = 0.9$ and $\tau_{med} = 0.75$).
\ENSURE Potential MPC Tasks $\mathcal{C}_{potential}$, Review Pool $\mathcal{C}_{review}$

\STATE Initialize $\mathcal{C}_{potential} \leftarrow \emptyset$, $\mathcal{C}_{review} \leftarrow \emptyset$

\FOR{each candidate $c \in \mathcal{C}_{raw}$}
    \STATE $diff, \ desc \leftarrow \mathrm{ExtractData}(c)$
    \STATE $prompt \leftarrow \mathrm{ConstructPrompt}(\mathcal{H}_{mpc}, diff, desc)$

    \COMMENT{Three-layer evaluation: 1.Primitive Check, 2.Protocol Logic, 3.Domain Specificity}
    \STATE $layer\_outputs, \ conf \leftarrow \mathcal{A}_{agent}.\mathrm{Evaluate}(prompt)$
    \STATE $c.\mathrm{metadata} \leftarrow layer\_outputs$ \COMMENT{Save analysis for completion-stage verification}

    \COMMENT{Autonomous Routing based on Confidence Score}
    \IF{$conf \ge \tau_{high}$}
        \STATE $\mathcal{C}_{potential} \leftarrow \mathcal{C}_{potential} \cup \{c\}$ \COMMENT{Automatically accepted}
    \ELSIF{$\tau_{med} \le conf < \tau_{high}$}
        \STATE $\mathcal{C}_{review} \leftarrow \mathcal{C}_{review} \cup \{c\}$ \COMMENT{Flagged for human adjudication}
    \ELSE
        \STATE Discard $c$ as Noise \COMMENT{Low Confidence or None}
    \ENDIF
\ENDFOR

\RETURN $\mathcal{C}_{potential}, \mathcal{C}_{review}$
\end{algorithmic}
\end{algorithm}

\section{Example SAST Rule and Flagged Patch} \label{app:sast_example}

To make the Domain-Specific SAST stream of Section~\ref{sec:verification_framework} concrete, Figure~\ref{fig:sast_example} pairs one of our context-aware Semgrep rules with an actual LLM-generated patch that the rule flags.

\begin{figure}[H]
    \centering
    \begin{minipage}[t]{0.48\textwidth}
        \textbf{A. Context-Aware SAST Rule (Semgrep YAML)}
        \begin{lstlisting}[language=YAML]
- id: mpc-reveal-in-loop
  patterns:
    - pattern-inside: |
        for ... in ...:
            ...
    - pattern-either:
        - pattern: $X.reveal()
        - pattern: $X.get_plain_text()
  message: "Unsafe Reveal: reveal() inside loop leaks secrets"
  severity: ERROR
        \end{lstlisting}
    \end{minipage}\hfill
    \begin{minipage}[t]{0.48\textwidth}
        \textbf{B. Flagged LLM Patch (CrypTen)}
        \begin{lstlisting}[language=Python]
def decrypt_model(self):
    for name, param in self.named_parameters():
        self.set_parameter(
            name,
            # FLAGGED: reveals secret inside loop
            param.get_plain_text()
        )
        \end{lstlisting}
    \end{minipage}
    \caption{Example of domain-specific cryptographic static analysis. (A) A customised Semgrep rule designed to detect the structural anti-pattern of decrypting secrets within a loop. (B) An actual LLM-generated patch for CrypTen demonstrating the vulnerability caught by this rule, which successfully passed functional tests but critically leaks intermediate parameters.}
    \label{fig:sast_example}
\end{figure}

By structurally isolating variables that retain secure states, our context-aware matching paradigm successfully mitigates false positives while precisely exposing logical security liabilities generated by LLMs.

\section{Dataset Difficulty Analysis} \label{app:difficulty}

\begin{figure}[H]
    \centering
    \includegraphics[width=1.0\textwidth]{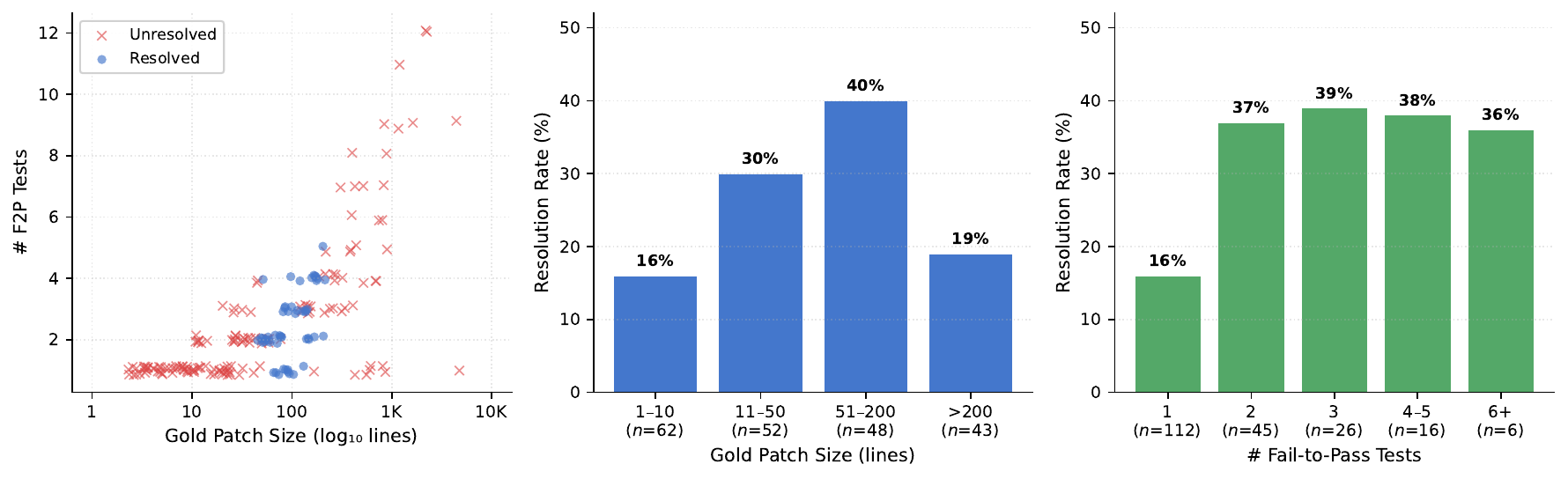}
    \caption{Difficulty analysis of MPC-Patch-Bench tasks. (Left) A scatter plot of Fail-to-Pass test count versus patch size for all instances, with resolved instances marked as circles~($\circ$) and unresolved as crosses~($\times$). (Center) Instance-level resolution rate grouped by gold patch size. (Right) Resolution rate grouped by number of F2P tests. Both axes reveal significant variation in difficulty across the benchmark.}
    \label{fig:difficulty}
\end{figure}

A primary motivation for MPC-Patch-Bench is to provide a benchmark that is genuinely challenging to state-of-the-art LLMs. Figure~\ref{fig:difficulty} presents a multi-faceted analysis of task difficulty as measured by resolution outcomes across 205 instances.

Tasks with gold patches in the 51--200 line range exhibit the highest resolution rate (40\%, $n=48$), suggesting that mid-complexity fixes---those large enough to require meaningful reasoning but small enough to fit within a model's effective context---represent the ``sweet spot'' where current LLMs demonstrate partial competence. In contrast, tasks requiring patches in excess of 200 lines drop to a resolution rate of only 19\% ($n=43$), and very short patches (1--10 lines) yield only 16\% resolution ($n=62$), the latter likely reflecting the need for highly precise, domain-critical one-liner edits that demand expert-level cryptographic understanding.

Analysing resolution by F2P test count reveals a complementary pattern: instances with 2 or more tests exhibit notably higher resolution rates (36--39\%), while those with a single F2P test yield only 16\% ($n=113$). We hypothesise that a higher test count provides richer feedback signals and clearer task boundaries that assist models in solution generation.

\newpage

\end{document}